\documentclass[twocolumn,showpacs,preprintnumbers,amsmath,amssymb,aps]{revtex4}
\usepackage{graphicx}% Include figure files
\usepackage{dcolumn}% Align table columns on decimal point
\usepackage{bm}% bold math

\begin{document}

\preprint{APS/123-QED}

\title{Saturation effects in the sub-Doppler spectroscopy of Cesium vapor \\ confined in an Extremely Thin Cell}% Force line breaks with \\

\author{C. Andreeva}
 \author{S. Cartaleva}
  \email{stefka-c@ie.bas.bg}
  \author{L. Petrov}
\affiliation{%
Institute of Electronics, Bulgarian Academy of Sciences, 72
Tzarigradsko Shosse boulevard, 1784 Sofia, Bulgaria
}%

\author{S. M. Saltiel}%
\affiliation{Sofia University, Faculty of Physics, 5 J. Bourchier
boulevard, 1164 Sofia, Bulgaria
}%

\author{ D. Sarkisyan, T.Varzhapetyan }%

 \affiliation{Institute for Physical Research, NAS of Armenia, Ashtarak-2,
Armenia
}%

\author{ D. Bloch, M. Ducloy  }
\affiliation{%
Laboratoire de Physique des Lasers \textbf{UMR 7538 du CNRS},
 Universit\'{e} Paris-\textbf{13}, F-93430 Villetaneuse, France}%

\date{\today}% It is always \today, today,

\begin{abstract}
Saturation effects affecting absorption and fluorescence spectra
of an atomic vapor confined in an Extremely Thin Cell (cell
thickness $L < 1 \mu m$) are investigated experimentally and
theoretically. The study is performed on the $D_{2}$ line
($\lambda ~= ~852 nm)$ of $Cs$ and concentrates on the two
situations $L = \lambda /2$ and $L =\lambda$, the most contrasted
ones with respect to the length dependence of the coherent Dicke
narrowing. For $L = \lambda /2$, the Dicke-narrowed absorption
profile simply broadens and saturates in amplitude when increasing
the light intensity, while for $L =\lambda$, sub-Doppler dips of
reduced absorption at line-center appear on the broad absorption
profile. For a fluorescence detection at $L =\lambda$,  saturation
induces narrow dips, but only for hyperfine components undergoing
a population loss through optical pumping. These experimental
results are interpreted with the help of the various existing
models, and are compared with numerical calculations based upon a
two-level modelling that considers both a closed and an open
system.
\end{abstract}

\pacs{42.50.Ct, 42.50.Gy, 42.62.Fi}% PACS, the Physics and Astronomy
\maketitle

\section{\label{sec:level1}Introduction }
High resolution spectroscopy of atoms confined in a thin cell is
promising for the investigation of complex spectra of atoms and
molecules. When irradiating a $10~-~1000\mu m$ thin cell under
normal incidence, the transmission spectrum of a single light beam
had revealed tiny  sub-Doppler features, nevertheless allowing the
resolution of the hyperfine  structure (h.f.s.) of alkali atoms
resonances \cite {1,2,3}. Indeed, the atoms travel wall-to-wall
and resonantly interact with light according to a transient
regime, yielding an enhanced response for atoms with longer
interaction time, i.e. for those atoms insensitive to a Doppler
shift. Recently, the extension to a technology of Extremely Thin
Cell (ETC, thickness $L < 1 \mu m$) \cite {4} has shown that these
sub-Doppler resonances become easily observed and can be a
dominant effect, even in the absence of a frequency modulation
(FM) technique. With these ETCs, an even narrower fluorescence
spectrum can also be detected, showing that transmission and
fluorescence behave differently. For the transmission spectrum, it
was shown that one observes a maximal coherent sub-Doppler Dicke
narrowing \cite {5} for a cell length $L = \lambda /2$, with
periodic revivals at $(2n+1)\lambda /2$ \cite {2,6}, while for $L
=\lambda$  (or $n\lambda$ ), the local coherent atomic response is
phase-mismatched all over the ETC, making the sub-Doppler features
vanishing and only allowing a residual Doppler-broadened response.
This length-dependent narrowing has no equivalent in the detection
of an incoherent process such as the fluorescence \cite {7}, whose
narrow features simply originate from the enhancement of the
contribution of those atoms whose velocity allows for a long
interaction time. This leads to a behavior that is monotonic with
the cell thickness. These features of sub-Doppler transmission and
fluorescence in ETCs were demonstrated in the linear regime of
interaction (with respect to intensity), allowing for example the
experimental determination of transition probabilities of the
respective h.f.s. components \cite {8}.

The detailed mechanisms for these sub-Doppler features are known
to be complex because, depending upon the irradiating intensity, a
coherent linear regime \cite {2} and an (incoherent) non-linear
regime of optical pumping had already been distinguished for the
absorption in relatively long cells ($L \geq 10\mu m$)\cite
{1,2,3}. As long as the atomic system can be described in a
two-level frame, various treatments for transmission experiments
were developed in an asymptotic regime relatively to the
irradiating intensity \cite {1,2,3,9}, up to a full analytical
formal expansion (\cite {10}). For ETCs, these high-intensity
effects become even more complex, with respect to the interplay
between the interferometric dependence associated to the Dicke
narrowing, and the velocity-dependent efficiency of saturation
mechanisms.

In this communication, we present an experimental and theoretical
study of the intensity effects in an ETC. For simplicity, we
restrict our experiments to the two cases $L = \lambda /2$ and $L
=\lambda$, the most important ones with respect to the periodicity
of the Dicke coherent narrowing. An additional simplicity for this
restriction is that it allows to neglect some of the Fabry-Perot
effects intrinsic to ETCs, namely the mixing of transmission with
reflection signals \cite {11} because the non resonant reflection
vanishes for these cell lengths. The study is performed on the
$D_{2}$ line of $Cs$ vapor, with a spectral resolution about an
order of magnitude better than in \cite {7}, and with an
irradiation intensity orders of magnitude larger than in previous
experiments on $Cs$ vapor in an ETC \cite {6,7}. In addition to
the known occurrence of narrow dips over a broader background in
the absorption spectrum \cite {7,12}, we discuss here the
appearance of narrow dips over the (sub-Doppler) fluorescence
spectrum. Such dips of reduced fluorescence were only briefly
described in preliminary reports \cite {12}. Here we present
experimental and theoretical studies aimed at the clarification of
the origin of the observed narrow dip in the fluorescence profile.
An interesting peculiarity of the observed narrow
reduced-fluorescence dip is that it appears for all h.f.s.
components but the one that does not suffer population loss due to
hyperfine and Zeeman optical pumping.

In spite of the complexity of the $Cs$ atomic system with respect
to saturation effects (as due to the many hyperfine and Zeeman
substates), we  show, numerically as well as on the basis of
general theoretical arguments, that the major features of our
observations can be interpreted in the frame of a two-level model
provided that closed and open atomic systems are distinguished.

\section{\label{sec:level2} Experimental set up }

\begin{figure}[h]
\centering
\includegraphics[height=6.5cm, width=8.5cm, scale=0.5]{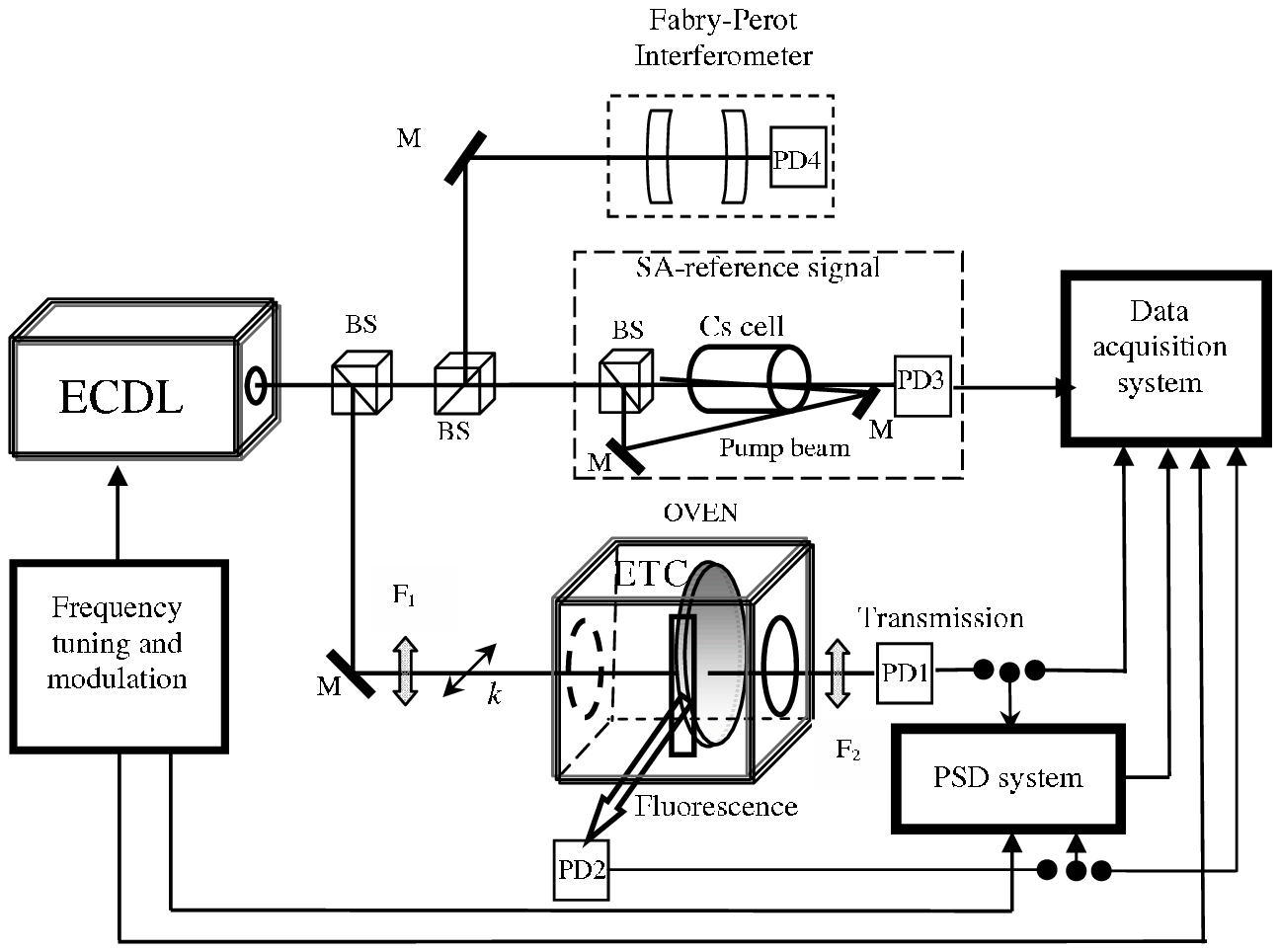}
\caption{ Experimental set up. }
 \label{Fig.1}
\end{figure}

A scheme of the experimental setup is presented in
Fig.\ref{Fig.1}. An extended cavity diode laser (ECDL) is used,
performing frequency-tunable single-mode operation at $\lambda ~=
~852nm$, with a FWHM of about $3MHz$. The main part of the laser
beam, linearly polarized, is directed at normal incidence onto the
ETC. The geometry of the experiment is chosen in a way that the
laboratory magnetic field (about $0.5G$) is approximately parallel
to the laser light polarization. The construction of the ETC,
filled with $Cs$ vapor from a side arm, is similar to the one
described in \cite {4}. Its design was slightly modified to
produce a wedge in the vapor gap. This  makes the cell thickness
locally variable in a convenient manner. The situations $L =
\lambda /2$ or $L =\lambda$ are chosen by simply adjusting the
relative position of the laser beam and of the ETC. The accuracy
of the cell thickness measurement is better than $20 nm$. The $Cs$
vapor density ($\sim 4.10^{13} at.cm^{-3}$) is controlled by the
temperature $T$ of the side arm (unless stated otherwise,
$T=119^{o}C$). The irradiating beam has a diameter of $0.4 mm$.
Its intensity is controlled with neutral density filters $F_{1}$.
The transmitted light power is measured by the photodiode $PD1$.
To ensure a constant sensitivity of the detector, the
off-resonance intensity falling onto $PD1$ is kept constant by
filters $F_{2}$. To record fluorescence spectra, the photodiode
$PD2$ collects the induced fluorescence emitted in a direction
normal to laser beam. The spectra can be recorded either directly,
or through the demodulation (with a Phase-Sensitive Detection
-PSD-) of a FM applied to the laser. Auxiliary laser beams allow
the monitoring of the laser frequency: (i) one beam is sent to a
scanning Fabry-Perot interferometer to monitor (by means of $PD4$)
the single-mode operation of the ECDL; (ii) the second one is used
for an auxiliary saturated absorption (SA) set-up with a
macroscopic (3-$cm$ long) $Cs$ cell ensuring an accurate reference
when scanning the ECDL frequency.

\section{\label{sec:level3}  Experimental results}

\subsection{Sub-Doppler resonances in absorption}

The absorption spectra comprise two sets of h.f.s. components (a
component being defined as optical transition between hyperfine
sub-states) $ F_{g} = 3\rightarrow F_{e} = \{2,3,4\}$ and $ F_{g}
= 4\rightarrow F_{e} = \{3,4,5\}$. They are represented in all the
following figures through the $\Delta P/P_{0}$ ratio (denoted as
absorption), where $\Delta P$ is the absorbed power, and $P_{0}$
is the input power. The relative uncertainty on the transmission
$(P_{0}- \Delta P)/P_{0}$ is on the order of a few $10^{-3}$.

\begin{figure}[h]
\centering
\includegraphics*[height=5.5cm, width=7.5cm, scale=0.5]{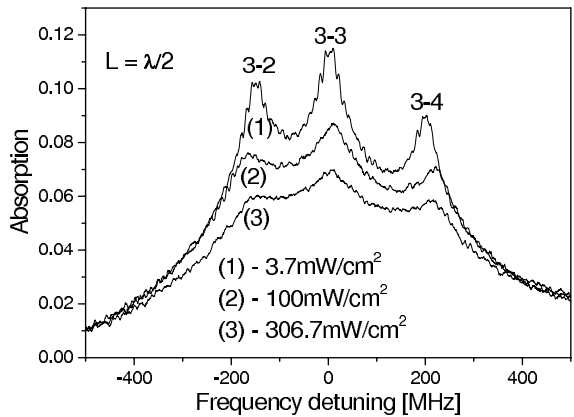}
\caption{Absorption spectra across the $ F_{g} = 3\rightarrow
F_{e} = \{2,3,4\}$ transitions for various intensities (as
indicated) at $L = \lambda /2$.}
 \label{Fig.2}
\end{figure}

Figure \ref{Fig.2} illustrates the behavior of the absorption
spectrum on a cell of a thickness $L = \lambda /2$, for three
different irradiating intensities and for the  $ F_{g} =
3\rightarrow F_{e} = \{2,3,4\}$ set of transitions (a similar
behavior is observed for the $ F_{g} = 4\rightarrow F_{e} =
\{3,4,5\}$ transitions). For all intensities, the enhancement of
the absorption at the center of the hyperfine transitions is
responsible for a strong narrowing of the spectrum, notably
allowing the resolution of the individual h.f.s. components. These
results extend those of \cite {6,7,12}, evidencing the coherent
Dicke narrowing; they are however obtained in an intensity range
higher than the one ($<< 1 mW/cm^{2}$) ensuring a genuine linear
behavior. At high intensities, the sub-Doppler resonances appear
significantly broadened, and saturation effects tend to washout
the Dicke coherent narrowing [13], which is well-pronounced at low
power: one can notice in Fig.\ref{Fig.2} that the absorption peaks
are strongly reduced for high intensities, but that the wings are
nearly unaffected.

\begin{figure}[h]
\centering
\includegraphics*[height=5.5cm, width=7.5cm, scale=0.5]{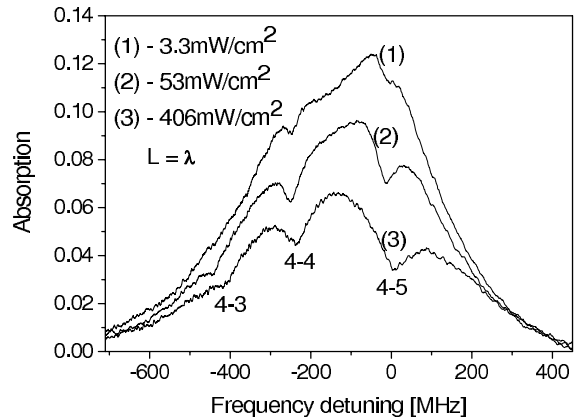}
\caption{Absorption spectra across the $ F_{g} = 4\rightarrow
F_{e} = \{3,4,5\}$ transitions for various intensities (as
indicated) at $L =\lambda$.}
 \label{Fig.3}
\end{figure}

Figure \ref{Fig.3} shows the typical evolution of the absorption
spectrum for $L =\lambda$ with the irradiating intensity (only the
$ F_{g} = 4\rightarrow F_{e} = \{3,4,5\}$ set is shown, but a
similar behavior is observed for the  $ F_{g} = 3\rightarrow F_{e}
= \{2,3,4\}$ transitions). Significant differences are observed
between the absorption spectrum at $L = \lambda /2$ and that at $L
=\lambda$. Let us first recall that for $L =\lambda$, no coherent
Dicke narrowing is expected, and that in the linear regime, the
absorption profile, although complex, is Doppler-broad, owing to a
(non velocity-selective) transient regime of interaction.
Superimposed to the expected Doppler-broadened absorption profile,
one observes, as a result of the relatively high intensities used
here, well pronounced sub-Doppler narrow dips of reduced
absorption. This reduction of absorption is a signature of optical
pumping and/or saturation processes that tend to reduce the number
of atoms available for the interaction with irradiating light.
These processes can be completed only for atoms interacting a
sufficient time with the laser light \cite {1,3}: they are highly
enhanced for slow atoms (i.e. small velocity component along the
normal to the ETC windows), hence yielding sub-Doppler structures.
The amplitude of these narrow structures increases when the
irradiating intensity increases, the structures get apparently
broader, and their contrast relative to the broad Doppler
absorption increases markedly, as the Doppler-broadened absorption
decreases under saturation. These results appear very similar to
those presented for the absorption spectra in ETC of $Rb$ vapor
\cite {12}, but include a regime of a higher irradiating
intensity.

\begin{figure}[h]
\centering
\includegraphics*[height=6cm, width=8cm, scale=0.5]{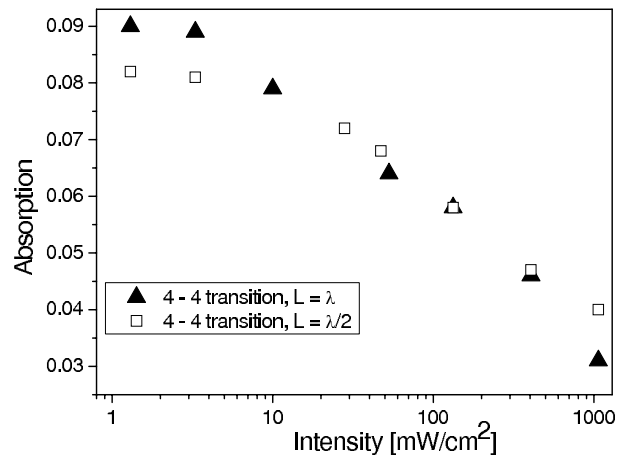}
\caption{Experimental absorption at the line center of the $F_{g}
= 4 \rightarrow F_{e} = 4 $ h.f.s. component, as a function of the
irradiating intensity for $L = \lambda /2$ and $L =\lambda$.}
 \label{Fig.4}
\end{figure}

Naturally, it is not a surprise that the narrow
(velocity-selective) saturation dips are observed more easily when
the non saturated lineshape is broad ($L =\lambda$), than when it
undergoes a notable coherent Dicke narrowing ($L = \lambda /2$).
To compare more quantitatively the differing saturation behaviors
for $L = \lambda /2$  and $L =\lambda$, the absorption at the
center of individual h.f.s. component is plotted in
Fig.\ref{Fig.4} as a function of the intensity. The comparison is
here restricted to the $F_{g} = 4 \rightarrow F_{e} = 4 $
transition, but similar results are obtained for the other h.f.s.
components. It can be seen that the absorption rate decreases
faster at $L =\lambda$ than at $L = \lambda /2$. This faster
reduction when the length increases could be seen as reminiscent
of the behavior of velocity-selective pumping already observed in
micrometric thin cells \cite {1}, when the efficiency of the
saturation process is governed by the product of the intensity by
the cell length. An additional discussion is provided in Section
IV.

\subsection{Narrow resonance in fluorescence}

\begin{figure}[h]
\centering
\includegraphics*[height=5.7cm, width=8cm, scale=0.5]{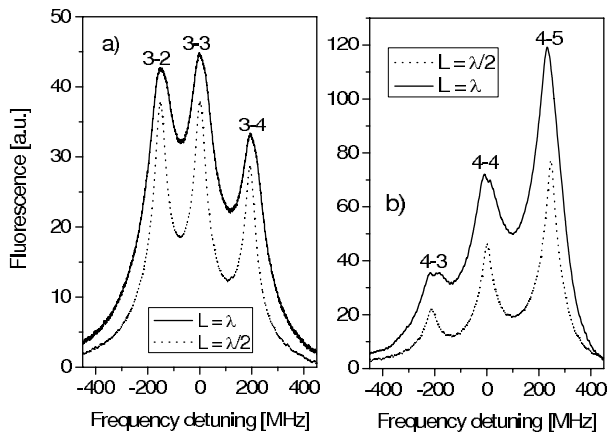}
\caption{ Illustration of the ETC fluorescence spectra for lower
(a, $7mW/cm^{2}$) and higher (b, $130mW/cm^{2}$) light intensities
at the two studied cell thicknesses. $Cs$ source temperature:
$127^{o}C$ (a) and $105^{o}C$ (b).}
 \label{Fig.5}
\end{figure}

\begin{figure}[h]
\centering
\includegraphics*[height=6cm, width=8cm, scale=0.5]{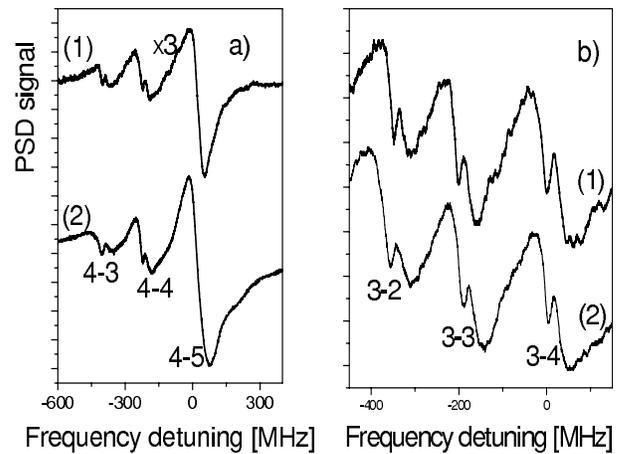}
\caption{ PSD fluorescence spectra (for $ L = \lambda $) as
obtained after demodulation of an applied FM on (a) the set $
F_{g} = 4\rightarrow F_{e} = \{3,4,5\}$, recorded with an
intensity : (1) -$140mW/cm^{2}$ ; (2) -$1076mW/cm^{2}$, and (b) on
the set $ F_{g} = 3\rightarrow F_{e} = \{2,3,4\}$ with an
intensity :  (1) -$306mW/cm^{2}$ ; (2) -$800mW/cm^{2}$ . In the
PSD spectra, the narrow dispersive structure exhibits a (quasi-)
antisymmetry opposed to the one of the broad structure. This is a
signature of a narrow dip in the corresponding spectrum for direct
detection. }
 \label{Fig.6}
\end{figure}

As previously reported \cite {4,7}, the fluorescence spectra
exhibit sub-Doppler features that are narrower than those in the
transmitted light, with an amplitude and width following a
monotonic growth with the cell thickness. Very well resolved
fluorescence spectra are recorded directly (Fig. \ref{Fig.5})
without FM and PSD of the signal. Even at low irradiation
intensities (Fig. \ref{Fig.5}a), the signals are narrower for  $L
= \lambda /2$ than for $L =\lambda$ . At high irradiating
intensities, we observe  for $L =\lambda$ tiny dips that are
superimposed to the top of the fluorescence profile of  the
individual h.f.s. components. Although these dips are observable
through a direct detection (Fig. \ref{Fig.5}b), they are more
conveniently characterized through the FM technique
(Fig.\ref{Fig.6}). Note as an additional difference between
fluorescence and absorption spectra, that these saturation dips
occur for much higher irradiating intensities ($\sim$ an order of
magnitude in our experiments) in fluorescence than in absorption.

To understand how these saturation features are specific to
fluorescence, two major differences with absorption are worth
being underlined : (i) while the absorption rate decreases to zero
under saturation effects (because the population difference is
reduced), increasing the irradiating intensity tends to increase
the population of the excited state, and hence the fluorescence
(at least as long the atomic system is a closed system); (ii) even
in the linear regime (i.e. low irradiating intensity), the
fluorescence is a second-order process \cite {4}, that is
velocity-selective in the transient regime specific to ETC. Hence,
the observation of narrow dips in an already narrow fluorescence
spectrum can seem intriguing: in particular, for a closed atomic
system, i.e. when no population loss occurs, the fluorescence rate
is expected to be maximal at line center once the steady-state
regime is reached. Moreover, one has to understand how a narrow
velocity-selective dip of population loss can be superimposed to
an already narrow velocity-selective peak of fluorescence. The
last point can be tempered by the fact that the selectivity of
atomic velocity decreases with increasing the cell length and is
responsible for a significant broadening for lengths exceeding
$\sim  \lambda /2$. This makes the width of a fluorescence
spectrum not as narrow as the pure natural width and this leaves
open the possibility of a more selective process (population loss,
assuming an open system) of the opposite sign; in addition, the
sub-Doppler fluorescence spectrum already undergoes a notable
broadening because of the high intensities required to observe
saturation dips.

Before further interpreting our experimental findings (see Section
IV), it is interesting to point out that for the $ F_{g} =
4\rightarrow F_{e} = \{3,4,5\}$ set of transitions, saturation
dips appear only for the open transitions $F_{g} = 4 \rightarrow
F_{e} = 3 $ and $F_{g} = 4 \rightarrow F_{e} = 4 $, but are not
observed for the closed transition $F_{g} = 4 \rightarrow F_{e} =
5 $ (Fig. \ref{Fig.6}a), in spite of the large explored range of
irradiation intensities $(50-1000 mW/cm^{2})$. Conversely, Fig.
\ref{Fig.6}b strikingly shows that for the $ F_{g} = 3\rightarrow
F_{e} = \{2,3,4\}$ set of h.f.s. components, all h.f.s.
components, including the closed transition $F_{g} = 3 \rightarrow
F_{e} = 2 $, exhibit a comparable saturation dip in the
fluorescence spectrum under the considered laser intensities.
Actually, it is known that due to the Zeeman degeneracy, a closed
transition pumped with polarized light cannot be simply viewed as
a transition on a (degenerate) 2-level system. It is in particular
not protected against the Zeeman optical pumping that modifies the
tensorial orientation of the hyperfine sub-level through the
excitation with linearly polarized light \cite {14}.  As a result
of a strong irradiation on the $F_{g} = 3 \rightarrow F_{e} = 2 $
transition, $Cs$ atoms accumulate into the  $m_{F}=\pm3$ Zeeman
sub-levels, which do not interact with the laser light: although
the transition is a closed one (in terms of energy level), a
strong irradiation  induces a decreased fluorescence, because the
system is actually an open one when considering the Zeeman
degeneracy. At the opposite, on the $F_{g} = 4 \rightarrow F_{e} =
5 $ transition, $Cs$ atoms accumulates on Zeeman sub-levels with
the largest absorption probability \cite {15,16}, and a strong
irradiation does not reduce the fluorescence.

%%%%%%%%%%%%%%%%%%%%%%%%%%%%%%%%%%%%%%%%%%%%%%%%%%%%%%%%%%%%%%%%%%%%%%%%%%

\section{\label{sec:level4} Discussion and interpretation with a theoretical modelling  }

\subsection{Limits in the interpretation and expectations from previous two-level models}

Although we describe a single-laser experiment, a fully
quantitative prediction, would be very complex to obtain. This is
because when dealing with saturation problems for degenerate
two-level system, the Rabi frequency of an elementary transition
between Zeeman components depends on geometrical Clebsh-Gordon
type coefficients, so that a single parameter of saturation can
hardly be defined.  Moreover, here, the saturation process is
governed by a transient regime, and the duration of the
interaction is velocity-dependent. This mixture of
velocity-dependent transient regime, and of the velocity
integration, justifies that several regimes have already been
analyzed in the elementary frame of a non degenerate two-level
model. In all cases, the theoretical treatment of spectroscopy in
a thin cell of dilute vapor assumes wall-to-wall atomic trajectory
\cite {1,2,3,9,13,17,18} and the optical response results from a
spatial integration of the local atomic response, that is
determined through a transient evolution.

For relatively long cells, an elementary treatment of saturation,
was developed \cite {1,3} (for an early independent approach, see
\cite {18}), relying basically on an open two-level model.
Saturation effects were considered to be much slower than the
coherent absorptive response, assumed to be instantaneous. They
induce a velocity-selective dip in the absorption spectrum. An
elementary scaling law was found, with the key parameter (the
pumping time) determined by the product "cell length by pumping
intensity". Even in this simplifying modelling, the identification
of the atomic velocities contributing to the signal \cite {3} has
revealed to be quite complex, because of an interplay between the
velocity width associated to the natural optical width, and the
maximal velocity allowing a quasi-steady state pumping.

For lower intensities and/or smaller cell length, the optical
pumping remains negligible, and an elementary (closed two-level)
model has to be considered \cite {2}. The relevant transient
regime is the build-up of the absorbing properties of the vapor,
i.e. of the optical coherence. The interference between the
various velocity-dependent (complex) coherent response of all
atomic velocities is at the origin of the Dicke narrowing at $L=
\lambda/2$ (and of its periodical revivals at $L = (2n+1)
\lambda/2)$. The Bloch vector model \cite {7} is an adequate tool
to explain the periodical Dicke narrowing for absorption and it
can be applied beyond the limits of the linear regime \cite
{19,20}, or to accommodate the population losses of an open
system. For a strongly driven irradiation, and a moderate
relaxation (closed system) , the global process remains purely
coherent, but requires the velocity integration (or interference)
of  quickly rotating Bloch vectors. Through the interference of
these multiple oscillations, it can be inferred \cite {19,20} that
varying the detuning, (i.e. changing the orientation of the
pseudo-magnetic field in the Bloch-vector model), will lead to an
oscillating behavior, instead of always yielding a maximum at line
center. And indeed, a formal analytical treatment \cite {10} for a
two-level model (closed or open system), predicts such a
multi-peaked absorption lineshape under a strong saturation,
starting with a simple dip at line center for moderate saturation.
However, in spite of its formal analyticity, this treatment
requires a numerical determination of the relevant eigenvalues
determining the solutions, and becomes cumbersome for a lineshape
calculation. Also, no specific analysis has been provided for the
situations the most relevant for the coherent Dicke narrowing,
only a situation closed to $L= 3 \lambda/2$ is explicitly studied.
In particular, for an open system, the problem of the competing
physics of the coherent response in the saturation regime, and of
the incoherent velocity-selective population depletion, has not
been addressed in \cite {10}. The systematic modelling of the
fluorescence response in the context of ETC has never been
reported. In \cite {13}, fluorescence spectra in ETCs are
calculated, but they mostly aim at the specific description of the
multi-level Rb transition.

As a further step of our study, we propose a numerical evaluation
of the response of an elementary non-degenerate two-level atomic
system (conservative or open 2-level system), and we compare it
with the experimental findings.

\subsection{A two-level modelling in view of a numerical estimate}

\begin{figure}[h]
\centering
\includegraphics*[height=4.5cm, width=8.5cm, scale=0.5]{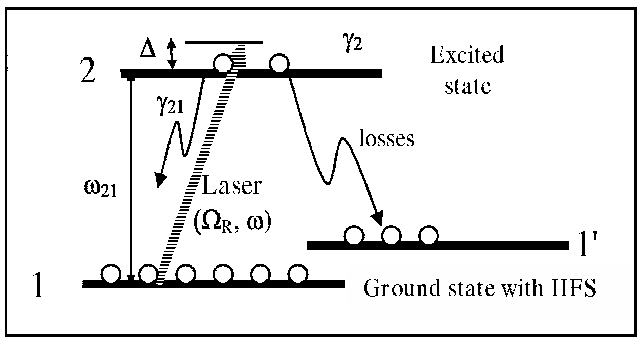}
\caption{ Schematics of the atomic system used in the theoretical
modelling.}
 \label{Fig.7}
\end{figure}

We consider here a two-level model with a control parameter
allowing to compare the situation of an open system (with losses
to a generic "third level" $1^{\prime}$), or of a closed system
(Fig.\ref{Fig.7}). Levels 1 and 2 are coupled by a laser light at
a frequency $\omega $, detuned by $\Delta $ from the transition
frequency $\omega_{21}$ ($\Delta = \omega - \omega_{21}$). The
Rabi frequency is defined as $ \Omega_{R} = 2\mu_{12}E_{0} /
\hbar$ ($ E_{0}$ being the light field input amplitude) with
$\mu_{12}$ the dipole moment of the transition. The width of the
excited state is denoted by $\gamma_{2}$, and one assumes, for the
considered dilute vapor, that collisions - notably dephasing
collisions- can be neglected, so the optical width of the
transition $\gamma_{21}$ is $ \gamma_{21} =  \gamma_{2} / 2$.
Practically, this simple hypothesis introduces a set of dual
relaxation constants. This unfortunately leads to hardly tractable
analytical solutions, while the assumption of a single relaxation
constant for population and optical coherence, would have greatly
simplified the calculations. However, such a simplifying
assumption can hardly be justified in the context of an ETC, with
the wall-to-wall atomic trajectories \cite {21}. To take into
account the possibility of population losses to a third level
(e.g. to the other hyperfine sub-level of the ground state of
alkali-metal atoms, or to the Zeeman sub-levels), one introduces a
coefficient, defined as $\alpha$, for characterizing the
probability to decay from level 2 to level 1 ($\alpha = 1$ for a
closed system, $0 \leq \alpha< 1$ for an open system).

The system of Bloch equations is hence the following:

\begin{equation}\label{Eq.(1)}
 v\frac{d\sigma_{21}}{dz} +D_{21}\sigma_{21}
-i\frac{\Omega_{R}}{2} \left( {\sigma_{11} -\sigma _{22} }
\right)= 0,
\end{equation}

\begin{equation}\label{Eq.(2)}
v\frac{d\sigma_{22}}{dz} +\gamma_{2} \sigma_{22} -\Omega_{R}Im
\sigma_{21} = 0,
\end{equation}

\begin{equation}\label{Eq.(3)}
v\frac{d\sigma_{11}}{dz} -\alpha \gamma_{2}\sigma_{22}
+\Omega_{R}Im\sigma_{21} =  0,
\end{equation}
where $ D_{21}=\gamma_{21}+ikv-i\Delta $ , $v$ is the atomic
velocity (along the laser beam, and hence along the normal to the
ETC), and the $\sigma_{ij}$ are the reduced density matrix
elements in the rotating frame. The above system had been solved
analytically \cite {1}, in an approach focusing only on the
non-linear incoherent processes. This could be justified in view
of solving the restricted problem for cell lengths allowing
transient coherent processes to be negligible. Here we consider
both coherent and incoherent processes that are altogether
essential for the sub-micron sized cells.

As usual in a thin cell, the initial conditions for the system of
Eqs(\ref{Eq.(1)}-\ref{Eq.(3)}) differ for arriving $(v < 0)$ and
departing $(v > 0)$ atoms. One has indeed $[{ \sigma_{11}(L) = 1;
\sigma_{22}(L) = 0; \sigma_{21}(L) = 0}]$ for $v < 0$; and $[{
\sigma_{11}(0) = 1; \sigma_{22}(0) = 0; \sigma_{21}(0) = 0}]$ for
$v > 0$. To relate the solution of the system (1-3) to the signals
of absorption or fluorescence observed in the experiments, we
further follow the approach presented in \cite {9}. The local
atomic response (at $z$) is deduced from the integration of its
transient behavior (owing to $z= vt$, or $z = L+vt$,  for
respectively $v > 0$ and $v < 0$). The optical signal results from
the spatial integration of the atomic response, after the required
integration over the velocity distribution (assumed to be a
Maxwellian, with a thermal velocity $u$).
 Hence, the absorption is proportional to a quantity

 \begin{equation}\label{Eq.(4)}
A=\int_{0}^{\infty} G(v)exp[-(\frac{kv}{ku})^{2}]dv
\end{equation}

with
\begin{equation}\label{Eq.(5)}
G(v)=\int_{0}^{L} Im[\sigma_{21}(z,v)]dz.
\end{equation}

In the experiment, the measured signal is the coherent beating
between the input field and the reemitted field $ I_{t}\sim
2E_{0}E_{t}^{\ast}$, so that it is the experimental ratio of
absorption
 $\triangle P/P_{0}$ which has to be compared with the theoretical quantity $
 A/\Omega_{R}$
(or $I_{t}/P_{0}$). In a similar way,  the fluorescence is related
to the quantity $U$, with :

\begin{equation}\label{Eq.(6)}
U=\int_{0}^{\infty} Q(v)exp[-(\frac{kv}{ku})^{2}]dv
\end{equation}

where $Q(v)$ is defined as

\begin{equation}\label{Eq.(7)}
Q(v)=\int_{0}^{L}[\sigma_{22}(z,v)]dz.
\end{equation}

On this theoretical basis, it is possible to spatially- and
velocity- integrate the solutions of the density matrix equations
to be found numerically under the conditions of saturation.
Relatively to the previous calculations \cite {1,2,3}, one should
recall that: (i) the modelling of the coherent Dicke narrowing was
achieved on the basis of similar density matrix equations \cite
{2,17}, in the limit of a first-order interaction with the
resonant light (in such a case, the parameter  $\alpha$, affecting
population redistribution, plays no role in this first order
prediction, and the Bloch vector model applies), and that (ii) for
the velocity-selective optical pumping in thin cells ($L
>> \lambda$ ) \cite {1}, the optical coherence yielding the absorption
rate was estimated under a rate equation approach, allowing the
instantaneous measurement of the remaining active population
difference (for $\alpha\leq1$).

In view of discussing some of the theoretical predictions with
parameters applicable to an elementary and realistic case, the
above model has been used with the following parameters:
$\gamma_{21}=5MHz$, $ku=250 MHz$, and  $\alpha=1$ (closed system)
or $\alpha=0.5$ for a realistic open alkali system. Technically,
our numerical results combine a velocity integration, the spatial
integration of a locally-varying response (see Eq.\ref{Eq.(5)} and
Eq.\ref{Eq.(7)}), and a Runge-Kutta integration equivalent to the
integration of the transient response governing the spatial
response for a given velocity.

\subsection{ Comparison between the model and the experiments : Absorption behavior }

\begin{figure}[h]
\centering
\includegraphics*[height=5.5cm, width=8.5cm, scale=1]{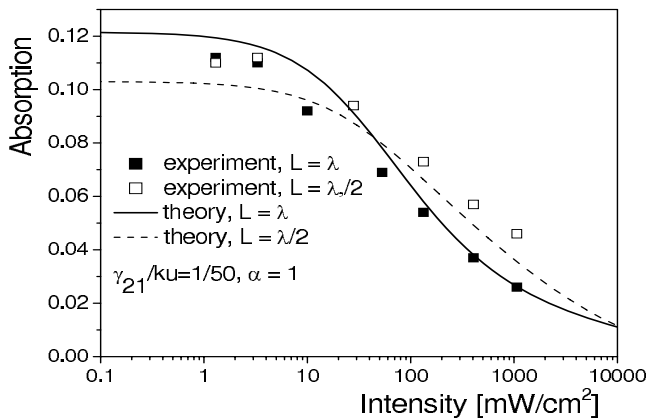}
\caption{Intensity dependence of the absorption - at line-center-
for a closed transition. The theoretical dependence for $ L =
\lambda /2$ is in dashed line, the solid line is for $ L = \lambda
$. The experimental data ($ L = \lambda /2$ : open squares, $ L =
\lambda $ : full squares) are for the closed $F_{g} = 4
\rightarrow F_{e} = 5 $ transition. For the scale applied to the
theoretical curves, see text. }
 \label{Fig.8}
\end{figure}

When attempting to compare quantitatively the experimental results
with a modelling, and especially if it is not intended to go to a
complete lineshape analysis, it is necessary to recall  various
intrinsic limitations affecting the possibility of a quantitative
comparison between the experiment and the above two-level model.
First, this comparison remains in principle of a limited scope
because multiple Zeeman transitions are involved. This means that
in principle, saturation effects cannot be accounted by converting
the experimental intensity into a single Rabi frequency
$\Omega_{R}$. The tensorial structure of the atomic system makes
non identical the various transfer rates to the individual
sub-levels. On the experimental side, the hyperfine components are
not perfectly resolved, but partially overlap, moreover in a non
constant manner that depends upon the cell length, and the
saturation. This makes uneasy to attribute all of the measured
absorption at a given frequency - on the center of a h.f.s.
component, or elsewhere - to a single h.f.s. component. This
limitation is even stronger for the smallest components because
they are observable only over the slope of a stronger component,
adding an extra-difficulty to characterize the appearance of an
inverted dip structure. A rigorous measurement for a given
hyperfine component would imply to subtract the contribution from
the neighboring components. Such an evaluation cannot be very
precise, and it becomes natural to concentrate the study on the
stronger transitions. One can also mention that the uncorrected
transverse structure of the irradiating intensity (presumably
Gaussian) tends to wash out the tiny oscillations that could be
induced by saturation \cite {10}, and that the residual
terrestrial magnetic field, although not sufficient to generate a
resolved Zeeman structure, may modify the coupling rates between
sub-levels.

\begin{figure}[h]
\centering
\includegraphics*[height=9cm, width=7cm, scale=1]{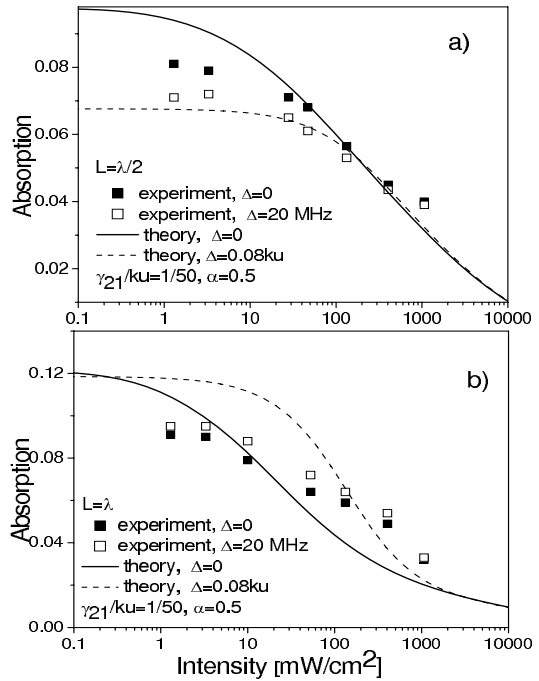}
\caption{Comparison between the intensity dependence of the
absorption on a open transition, at line-center, and at a slightly
detuned frequency. The ETC thickness is (a) $ L = \lambda/2 $; (b)
$ L = \lambda $ . The theoretical dependence is shown in solid
line for a frequency at line center $\Delta=0$, and in dashed line
for a detuned frequency   $\Delta=0.08 ku$. The experimental data
($ \Delta=+20 MHz$ : open squares, $ \Delta = 0$ : full squares)
are for the open $F_{g} = 4 \rightarrow F_{e} = 4 $ transition.
For the scale applied to the theoretical curves, see text. }
 \label{Fig.9}
\end{figure}

\begin{figure}[h]
\centering
\includegraphics*[height=9.5cm, width=7.5cm, scale=1]{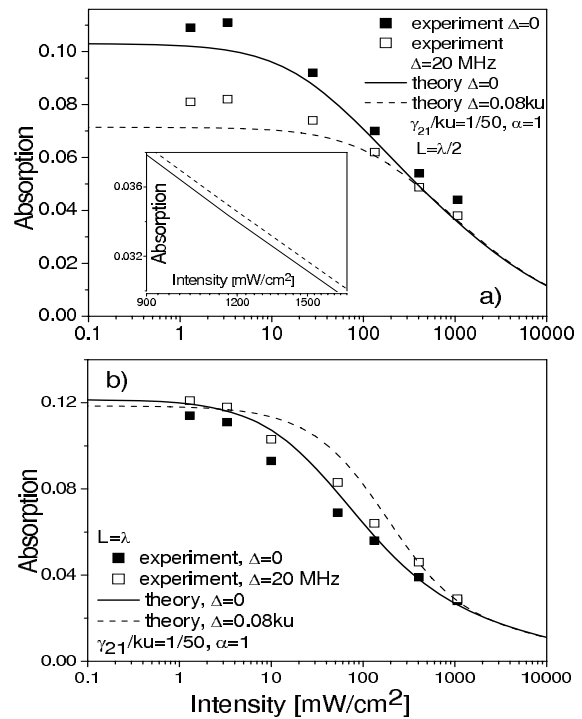}
\caption{Same as Fig. \ref{Fig.9}, but for a closed transition
theory (experimental data from the  $ F_{g} = 4 \rightarrow F_{e}
= 5$ transition). }
 \label{Fig.10}
\end{figure}

For all the above reasons, the interplay between numerous
processes, with differing time constants, makes hopeless the
characterization of the complex broadening of lineshapes by a
"width" of resonance. This is why, in an attempt to simply
evaluate the onset of  the appearance of a narrow inverted
structure (i.e. reduced absorption), we compare predictions for an
irradiation frequency at line-center, and for a slightly shifted
frequency (we take $\Delta = 0.08 ku$,  or $20 MHz$ for numerical
values as mentioned in Section IV-B). Such a criterion, possibly
misleading if the spectrum would include numerous oscillations,
seems reasonable with respect to the apparent width of the various
saturation dips that we observe. Figs.\ref{Fig.8}-\ref{Fig.10}
allow a comparison between the experimental and the theoretical
results for the $F_{g} = 4 \rightarrow F_{e} = 4 $ and $F_{g} = 4
\rightarrow F_{e} = 5 $ transitions. These two transitions are
good examples of open and closed transitions, and the
corresponding "saturating intensity" (although the concept is, as
mentioned, of a limited scope for a degenerate system) should be
quite comparable. To make the theoretical predictions directly
comparable to the experiments, we use a conversion factor
$(\Omega_{R}/ \gamma_{21})^{2} =1$ for $15mW/cm^{2}$ which was
chosen to provide the most satisfactory visual fit between the
experimental and theoretical curves. Also, in these figures, the
absorption rate for the theoretical curves was adjusted (by a
factor of 1.8) to provide the optimal comparison with the
experiments; note however that in principle, the absorption rate
is predictable in an absolute manner provided that the atomic
density and the dipole moment are known.

This comparison between the simplified modelling and the
experimental observations shows a satisfactory agreement. In
particular, in Fig. \ref{Fig.8}, where the predicted absorption at
line center is plotted in the two typical cases $ L = \lambda /2
$, and $ L =\lambda $  , one notes as predicted that if the
absorption at $ L = \lambda /2 $ is only slightly smaller than for
$ L =\lambda $   for low intensities (as expected due to the Dicke
narrowing, the exact ratio being governed by the $\gamma_{21}/ku$
factor), the absorption becomes even larger for $ L = \lambda /2 $
than for $ L =\lambda $  at higher intensities. Interestingly,
such a result is valid for a closed system (Fig. \ref{Fig.8}) as
well as for an open system - see Fig.\ref{Fig.9}. Although the
saturation processes for $ L =\lambda $  that reduce absorption of
slow atoms are in principle twice more efficient than the
comparable processes for $ L = \lambda /2 $ , the dominant effect
seems here to be the survival of the coherent Dicke narrowing (for
$ L = \lambda /2 $ ), with its large contribution of fast atoms
that are nearly insensitive to the  saturation. This larger
contribution at line center is the distinctive evidence of the
coherent Dicke narrowing, induced by the coherent transient
contribution of atoms that are not "slow". It is hence natural
that the Dicke coherent narrowing remains quite robust, as
unaffected by relatively strong irradiating intensities. However,
it cannot be concluded that a narrow saturation dip in the
absorption would not be observed in the conditions allowing for a
revival of the Dicke narrowing, such as $ L =3 \lambda /2 $  (a
length unfortunately not attainable because of the construction of
our cell): indeed, in most experimental conditions, the revival of
the Dicke narrowing \cite {6,7} (although shown to be robust with
saturation at $ L =3 \lambda /2 $, see \cite {12}) only brings a
sub-Doppler structure of a small amplitude.

Figures \ref{Fig.9} and \ref{Fig.10} allow the comparison of the $
L = \lambda /2 $  and  $ L = \lambda $  situations, with respect
to the appearance of a narrow dip of reduced absorption, for
closed and open systems. For the closed system considered in
Fig.\ref{Fig.10}, one predicts (Fig.\ref{Fig.10}b) for $ L =
\lambda $ that the initially broad peak - i.e. no Dicke narrowing-
exhibits an inverted substructure even for low intensities (a few
$mW/cm^{2}$) . Conversely, for $ L = \lambda /2 $
(Fig.\ref{Fig.10}a), the narrow Dicke structure undergoes only a
visible broadening, but without the clear appearance of a dip in
the center of the transition. A closer look on the inset of Fig.
\ref{Fig.10}a shows however that  $\Delta=0$ is no longer the peak
of absorption for high intensity, but the amplitude of the
corresponding dip is predicted to be extremely small. This
demonstrates that the absence of observation of a narrow dip for $
L = \lambda /2 $  is not some fundamental effect, but rather the
quantitative result of the competition between distinct processes
affecting optical coherences (Dicke narrowing) or atomic
population (saturation). An analogous behavior is predicted  for
an open system (Fig.\ref{Fig.9}, $\alpha=0.5$), with saturation at
line center and a tiny dip for $ L = \lambda /2 $ at high
intensities (slightly more pronounced than for the closed system),
and the occurrence of a pronounced narrow dip for $ L = \lambda $.
For $ L = \lambda $, the dip amplitude is predicted to be
significantly larger for the open transition than for the closed
one which is not observed in the experiment. For these
discrepancies, it should however be kept in mind that our analysis
here tackles narrow details of the  lineshapes, and  that a  shift
of $20 MHz$ is experimentally small.

\subsection{Predictions for the fluorescence behavior}

Because saturation effects in fluorescence are observed for higher
intensities than in absorption, and mostly in a PSD technique
following an applied FM, we have not attempted to perform a
quantitative comparison between the experiments, and the
predictions of the modelling.

\begin{figure}[h]
\centering
\includegraphics*[height=6.5cm, width=7cm, scale=1]{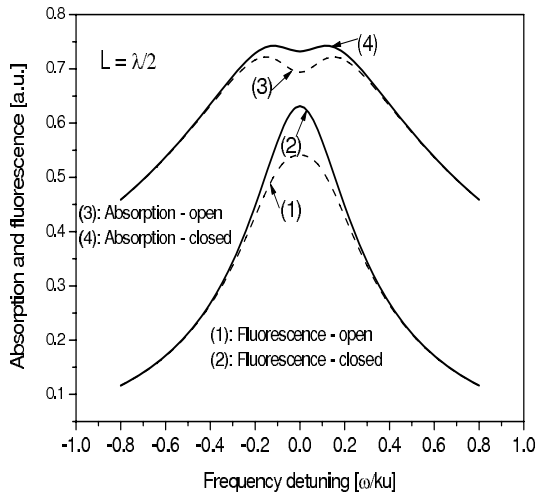}
\caption{Calculated fluorescence (1,2) and absorption (3,4)
profiles for open (1,3) and closed (2,4) transitions under
conditions where reduced absorption dip is predicted:
$\Omega_{R}/\gamma_{ 12} = 8$. }
 \label{Fig.11}
\end{figure}

The numerical calculations confirm the experimental observation
that lineshapes are expected to be narrower  in fluorescence than
in absorption in comparable conditions \cite {4,7,13}. Also, the
width of the fluorescence profile  is expected to increase
continuously with the cell thickness, without an interferometric
Dicke-type narrowing, and to reach a Doppler-broadened lineshape
for longer cells with the velocity selection getting less
stringent. The simulation in Fig.\ref{Fig.11} (i.e. calculation
with the relevant experimental parameters) does not predict the
formation of a narrow dip  in fluorescence for  $ L = \lambda /2 $
and this agrees with our experiment (Fig. \ref{Fig.5}), while
under the same condition, a dip in the absorption is predicted. If
such an absence of a dip can be expected for a closed system, a
strong irradiation should be able to induce a severe depletion of
the fluorescent atoms for an open system, and hence a dip in the
lineshape. However, in the sense of the dip formation, our
simulation does not show essential differences between the closed
and open transitions (Fig.\ref{Fig.11}). Most probably this is
because, for our choice of  $\gamma_{12}/ku$ parameter this strong
irradiation would imply for  $ L = \lambda /2 $ such a large
broadening of the transition that the velocity-selection itself is
governed by a width not markedly narrower than the one of the
total signal.

\begin{figure}[h]
\centering
\includegraphics*[height=12cm, width=7.5cm, scale=1]{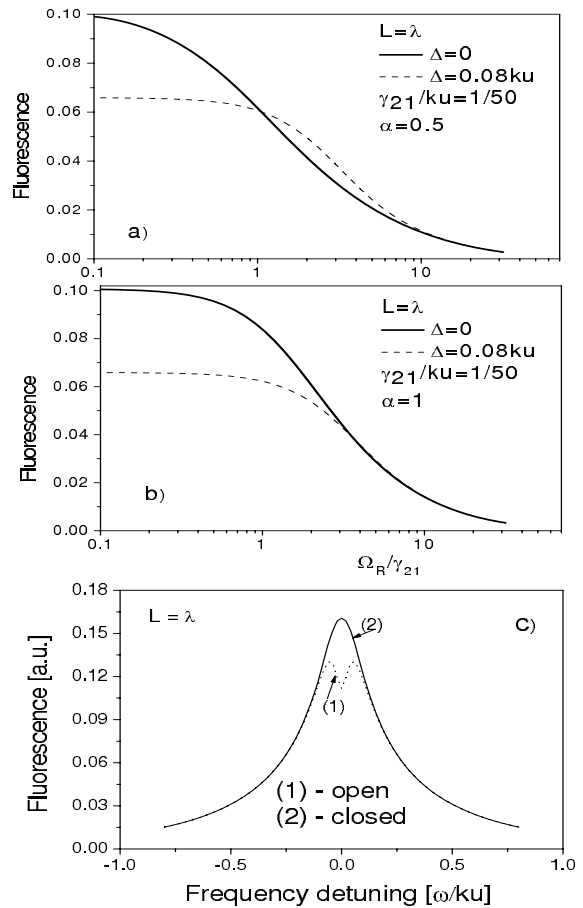}
\caption{Theoretical features of fluorescence for $ L = \lambda $.
Intensity dependence of the fluorescence (normalized by the Rabi
frequency, i.e. plot of $U/\Omega_{R}^{2}$ ) at line center
$\Delta=0$ (solid line), and at a slightly detuned frequency
$\Delta=0.08ku$ (dashed line) for : (a) an open transition; and
(b) a closed transition. (c) Theoretical lineshapes for open (1)
and closed (2) transitions for $\Omega_{R}/ \gamma_{21} = 2$. }
 \label{Fig.12}
\end{figure}

When increasing the cell length to $ L = \lambda $,
Fig.\ref{Fig.12}a,b shows, with a presentation similar to the one
used in Figs \ref{Fig.9}-\ref{Fig.10}, that a pronounced dip of
reduced fluorescence at line center is predicted at $ L = \lambda
$ for the open transition. For a closed transition, no dip is
predicted, and rather, the lineshape broadens with saturation. In
addition, theoretical fluorescence profiles are presented in
Fig.\ref{Fig.12}c, showing completely different behavior of the
fluorescence around the transition center for the open and closed
transitions. This striking theoretical difference between closed
and open system justifies our experimental observations (Sect.
III-B, and Fig.\ref{Fig.5}), where a fluorescence dip is observed
for all components but the $ F_{g} = 4 \rightarrow F_{e} = 5$
transition.

\section{\label{sec:level5} Conclusions }

In spite of the apparent simplicity of single beam experiments on
ETCs, we  demonstrate a large variety of regimes in the study of
saturation. This is  because  thin cell spectroscopy naturally
yields a signal averaged on various regimes of transient
interaction between an atomic velocity group, and  a resonant
irradiation. This also explains that a large variety of modelings
has been proposed to deal with these effects.  As long as the real
system of an alkali atom (such as $Cs$ in  our case) is far from
being a two-level system, owing to its degenerate multi-level
nature and including the tensorial structure responsible for the
various Zeeman sub-states, it is hopeless to describe in full
detail the saturation effects : this can be easily understood by
recalling that for alkali vapor, there exists no general
description of saturated absorption spectra under a strong pump
irradiation: this latter problem is however notably simpler as
being limited to a steady state interaction, but similarly
sensible to the many coupling strengths involved in the highly
complex sub-Doppler atomic structure of alkali atoms. In this
context, it becomes clear that an exact quantitative description
would be an enormous task of a probably limited benefit.
Nevertheless, it is remarkable that a comparison between a pure
two-level model, with well-chosen numerical parameters, and our
experiments, leads to a relatively satisfactory agreement.

On more general grounds, some major features can be deduced from
our studies. In absorption, the saturation reduces preferentially
the contribution of slow atoms. This leads to the observation of
saturation dips on line centers, that are observed more easily
when the (non saturated) absorption is broad (i.e. for $ L =
\lambda $ ), than when the coherent Dicke narrowing makes the
lineshape intrinsically narrow. For $ L = \lambda /2$, the Dicke
narrowing is so robust that we only observe a broadening, without
the occurrence of the predicted tiny saturation dip. Because of
the coherent nature of absorption processes in ETCs, the observed
dips at line centers can result from the combined
velocity-selective depletion of population difference, and from a
complex oscillating behavior. These oscillations are probably more
efficient for ETC thickness leading to a coherent narrow Dicke
structure, than for lengths multiple of $ L = \lambda $,
characterized by destructive interferences across the
Doppler-broad structure, and for closed systems rather than for
open systems (as characterized by an incoherent velocity-selective
population transfer).

In fluorescence, it is only for open atomic systems, allowing a
reduced contribution of slow atoms, that dips at line center can
be observed as originating from a velocity-selective process.
Also, increasing the cell length makes easier the observation of a
narrow structure inside the sub-Doppler fluorescence spectrum as
due to the increased width of the non-saturated fluorescence
spectrum. For effective closed systems (i.e. including the
redistribution among the Zeeman sub-levels), it is not clear if
the fluorescence spectrum can involve oscillations, reminiscent of
the kind of fringes that are predicted to appear  in the
absorption spectrum. This possibility could strongly depend upon
the relative relaxation of the optical coherence rate, and
population losses. In our experiments, the relatively strong
coherence losses, owing to the relatively high $Cs$ temperature,
and the uncontrolled spatial distribution of the irradiating beam,
could be sufficient reasons to make unobservable saturation
features more complex than an elementary dip.

The reported results enhance our knowledge in the rich field of
the Doppler-free ETC spectroscopy which is of significant
importance for the development of high-resolution spectroscopy of
atoms and molecules confined in nano-volumes. ETC  spectroscopy
has recently been shown to allow the spatial analysis of the
long-range van der Waals atom-surface attraction, that modifies
spectra for short ETC thicknesses \cite {22}. The recent
observation \cite {23} of Electromagnetically Induced Transparency
effect in  ETC is promising for the dynamics study of this widely
used phenomenon. ETC application has been proposed \cite {24} for
magnetic field measurements with sub-micrometer spatial resolution
which can be useful for detailed magnetic mapping performance.

\begin{acknowledgments}

The work is supported by the INTAS South-Caucasus Project (grant:
06-1000017-9001),  by the French-Bulgarian Rila collaboration
(French grant: 98013UK, Bulgarian grant: 3/10), by the National
Science Fund of Bulgaria (grant: F-1404/04)and enters into the
goal of the FASTnet consortium (EU support HPRN-CT-2002-00304). We
appreciate the help of K. Koynov with the numerical modelling. D.S
and T.V. would like to acknowledge ANSEF for the financial support
(grant: PS-nano-657).

\end{acknowledgments}

%\end{enumerate}


\begin{thebibliography}{999}%

%\begin{enumerate}
\bibitem {1} S. Briaudeau, D. Bloch, M. Ducloy, Europhys. Lett. \textbf{35}, 337, (1996).
\bibitem {2} S. Briaudeau, S. Saltiel, G. Nienhuis, D. Bloch, and M. Ducloy,  Phys. Rev. \textbf{A57}, R3169 (1998).
\bibitem {3} S. Briaudeau, D. Bloch, M. Ducloy, Phys. Rev. \textbf{A59}, 3723 (1999).
\bibitem {4} D. Sarkisyan, D. Bloch, A. Papoyan, M. Ducloy, Opt. Commun. \textbf{200}, 201 (2001).
\bibitem {5} R. H. Romer, R. H. Dicke, Phys. Rev. \textbf{99,} 532 (1955).
\bibitem {6} G. Dutier, A. Yarovitski, S. Saltiel, A. Papoyan, D. Sarkisyan, D. Bloch, M. Ducloy, Europhys.
Lett. \textbf{63},35, (2003).
\bibitem {7} D. Sarkisyan, T. Varzhapetyan, A. Sarkisyan, Yu. Malakyan, A. Papoyan, A. Lezama, D. Bloch, M. Ducloy,
 Phys. Rev. \textbf{A69}, 065802(2004).
\bibitem {8} D. Sarkisyan, T. Becker, A. Papoyan, P. Thoumany, H. Walther, Appl. Phys. \textbf{B76}, 625
(2003).
\bibitem {9} G. Dutier, S. Saltiel, D. Bloch, M. Ducloy, J. Opt. Soc. Am. \textbf{B 20}, 793
(2003).
\bibitem {10}H. Tajalli, S. Ahmadi, A. Ch. Izmailov, J. Opt. \textbf{B 4} , 208
(2002).
\bibitem {11}Note however that if the "reflection-like" (dispersive) component in the transmission
is eliminated for $L=n \lambda/2$ ( $n$ integer), the field
structure inside the ETC (see Ref. 9) is not a running plane wave;
rather, its amplitude exhibits a maximum at $z=(2n+1)\lambda /4$
that signs the (partially) standing wave nature of the field
inside the ETC. This non constant field amplitude is in the
principle susceptible to affect the saturation behavior, but the
spatial averaging intrinsic to the ETC response makes this effect
quite marginal.
\bibitem {12}T. Varzhapetyan, D. Sarkisyan, L. Petrov, C. Andreeva, D. Slavov,
 S. Saltiel, A. Markovski, S. Cartaleva, Proc. SPIE \textbf{5830}, 196 (2005) ; D. Sarkisyan,
 T. Varzhapetyan, A. Papoyan, D.Bloch, M.Ducloy, Proc.SPIE \textbf{6257}, 625701 (2006).
\bibitem {13}G. Nikogosyan, D. Sarkisyan, Yu. Malakyan, J. of Optical Technology \textbf{71}, 602 (2004).
\bibitem {14}Note that the Zeeman tensorial orientation of sub-level does not require an external magnetic
field, but is a signature of the nonisotropic nature of an atom
with angular momentum. Practically, residual magnetic fields are
not compensated in the experiment, and are susceptible to induce a
more intricate behavior, at least quantitatively, see ref. 1.
\bibitem {15}K. A. Nasyrov, Phys. Rev. \textbf{A63}, 043406 (2001).
\bibitem {16}F. Renzoni, C. Zimmermann, P. Verkerk, E. Arimondo,
J. Opt.B: Quantum Semiclassical Opt. \textbf{3}, S7 (2001).
\bibitem {17}B. Zambon, G. Nienhuis, Opt. Commun. \textbf{143}, 308 (1997);
see also  T. A. Vartanyan and D.L. Lin, Phys. Rev.
\textbf{A51},1959 (1995).
\bibitem {18}A. Ch. Izmailov, Laser Phys., \textbf{2}, 762(1992); \textbf{3},507 (1993);
Opt. Spectrosc., \textbf{74}, 25 (1993); \textbf{75}, 395 (1994).
\bibitem {19}I. Hamdi, P.Todorov, A. Yarovitski, G. Dutier, I. Maurin, S. Saltiel,
 Y. Li, A. Lezama, T. Varzhapetyan, D. Sarkisyan, M.-P. Gorza, M. Fichet, D. Bloch,
 M. Ducloy, Laser Physics \textbf{15}, 987(2005).
\bibitem {20}I. Maurin, P. Todorov, I. Hamdi, A. Yarovitski, G. Dutier, D. Sarkisyan,
 S. Saltiel, M.-P. Gorza, M. Fichet, D. Bloch and M. Ducloy, J. of Physics: Conference
 Series \textbf{19}, 20(2005).
\bibitem {21}Dephasing collisions do not necessarily affect atomic trajectories,
and the coherent Dicke narrowing has been demonstrated for various
atomic densities, i.e. with optical width possibly much larger
than the natural width, as defined by the half  of the inverse
lifetime of the excited state.
\bibitem {22} M. Fichet, G. Dutier, A. Yarovitsky, P. Todorov, I. Hamdi, I. Maurin,
S. Saltiel, D. Sarkisyan, M.-P. Gorza, D. Bloch, M. Ducloy,
Europhys. Lett. \textbf{77}, 540001 (2007)
\bibitem {23} A. Sargsyan, D. Sarkisyan, A. Papoyan,  Phys. Rev. \textbf{A73},
033803(2006).
\bibitem {24} D. Sarkisyan, A. Papoyan, T. Varzhapetyan, K. Blush,
M. Auzinsh, J. Opt. Soc. Am. \textbf{B 22}, 88 (2005).
\end{thebibliography}
\end{document}